\documentclass[10pt]{iopart}
\usepackage{iopams}
\usepackage{amssymb}
\usepackage{graphicx}
\usepackage{color}
\usepackage[tight,TABTOPCAP]{subfigure}
\usepackage{epstopdf}
\begin{document}
\makeatletter

%Feynman slash
\newbox\slashbox \setbox\slashbox=\hbox{$/$}
\newbox\Slashbox \setbox\Slashbox=\hbox{$/$}
\def\pFMslash#1{\setbox\@tempboxa=\hbox{$#1$}
  \@tempdima=0.5\wd\slashbox \advance\@tempdima 0.5\wd\@tempboxa
  \copy\slashbox \kern-\@tempdima \box\@tempboxa}
\def\pFMSlash#1{\setbox\@tempboxa=\hbox{$#1$}
  \@tempdima=0.5\wd\Slashbox \advance\@tempdima 0.5\wd\@tempboxa
  \copy\Slashbox \kern-\@tempdima \box\@tempboxa}
\def\FMslash{\protect\pFMslash}
\def\FMSlash{\protect\pFMSlash}
\def\miss#1{\ifmmode{/\mkern-11mu #1}\else{${/\mkern-11mu #1}$}\fi}
%%%% Uso:  \pFMSlash{p}
\makeatother

%%%%%%%%%%%%%%%%%%%%%%%%%%%%%%%%%%%%%%%%%%%%%%%%%%%%%%%%%%%%%%%%%

\title[Implications of Lorentz violation...]{Implications of Lorentz violation on Higgs-mediated lepton flavor violation}
\author{M. A. L\' opez--Osorio, E. Mart\'\i nez--Pascual, and J. J. Toscano}
\address{Facultad de Ciencias F\'{\i}sico Matem\'aticas,
Benem\'erita Universidad Aut\'onoma de Puebla, Apartado Postal
1152, Puebla, Puebla, M\'exico.}

\begin{abstract}
The lepton flavor violating decay of the Higgs boson $H\to l_Al_B$ is studied within two qualitatively different extensions of the Yukawa sector: one renormalizable and the other nonrenormalizable; both incorporating  Lorentz violation in a model-independent fashion.
These extensions are characterized by Yukawa-like matrices, the former by the constant Lorentz 2-tensor $Y^{AB}_{\mu \nu}$, whereas the latter by the constant Lorentz vector $Y^{AB}_\mu$.  It is found that the experimental constraints on the decays $l_A\to l_B\gamma$ severely restrict lepton flavor violating Higgs signals in the renormalizable scenario, as the electromagnetic transitions arise at tree level. In this context, it is found that the branching ratios of the decays $H\to \mu^\pm e^\mp$ and $H\to \tau^\pm \mu^\mp$ cannot be larger than $10^{-19}$ and $10^{-11}$, respectively. In the nonrenormalizable scenario, the electromagnetic transitions arise at one-loop level and transitions mediated by the Higgs or the $Z$ gauge boson are induced at tree level, hence we find mild restrictions on lepton flavor violation. Using the experimental limits on the three-body decays $l_A \to l_B \bar{l}_Cl_C$ to constraint the vector  $Y^{AB}_\mu$, it is found that the branching ratio for the decays $H\to \mu^\pm e^\mp$ is of about  $4\times 10^{-9}$, more important, a branching ratio of $7\times 10^{-4}$ is found for the $\tau^\pm \mu^\mp$ mode. Accordingly, the decay $H \to \tau^\pm \mu^\mp$ could be at the reach of future measurements. The lepton flavor violating decays of the $Z$ gauge boson were also studied. In the renormalizable scenario, it was found the undetectable branching ratios $BR(Z\to \mu^\pm e^\mp)<5.7\times 10^{-21}$ and $BR(Z\to \tau^\pm \mu^\mp)<2.0\times 10^{-12}$. As far as the nonrenormalizable scenario is concerned, it was found that $BR(Z\to \mu^\pm e^\mp)<0.67\times 10^{-12}$ and $BR(Z\to \tau^\pm \mu^\mp)<1.12\times 10^{-7}$. Although the latter branching ratio is relatively large, it still could not be within the range of future measurements.
\end{abstract}


\maketitle

\section{Introduction}
The flavor, an identity of certain elementary particles, is a feature of the Standard Model (SM) whose real origin still needs more exploration. On the other hand, in the SM, the recently discovered Higgs boson~\cite{ATLAS-CMS} is responsible for the masses of all known elementary particles. This field endows with different masses to quarks and leptons through couplings proportional to such masses; this peculiarity suggests that the Higgs boson is able to distinguish the flavor of each elementary particle. There are many open questions about the flavor; for instance, is it possible to find flavor transitions in the lepton sector at high energies, just like those already present in the quark sector? In this respect, the observation of the neutrino masses and mixing~\cite{PDG} marks the first evidence of flavor violation in the lepton sector. Since the absolute conservation of lepton flavor is considered a key aspect of the SM, the neutrino oscillation constitutes a first example of physics beyond the SM. These considerations motivate the study of lepton flavor violation (LFV) among charged leptons. Due to the peculiar role played by the Higgs boson concerning the flavor identity of particles, the phenomenon of lepton flavor violation mediated by this field deserves special attention.

In the SM, lepton flavor-changing neutral currents can be mediated by  the photon, the $Z$ weak gauge boson, and the Higgs boson. From these three options, the electromagnetic transitions $l_A\to l_B\gamma$, with $A\neq B$ and $l_{A}=e,\mu,\tau$, present the most stringent experimental constraints. Current bounds are given by~\cite{PDG}
\begin{eqnarray}
\label{BRgm1}
BR_{\,\textrm{Exp}}(\mu \to e\gamma)&& < 5.7\times 10^{-13} \ , \\
\label{BRgm2}
BR_{\,\textrm{Exp}}(\tau \to e\gamma)&& < 3.3\times 10^{-8} \ , \\
\label{BRgm3}
BR_{\,\textrm{Exp}}(\tau \to \mu \gamma)&& < 4.4\times 10^{-8} \ .
\end{eqnarray}
As far as the $Z$ gauge boson is concerned, the allowed LFV decays are less restricted by the experiment; the most recent bounds are given by~\cite{PDG}
\begin{eqnarray}
\label{BRZ1}
BR_{\,\textrm{Exp}}(Z\to e^\pm \mu^\mp) & &< 7.5 \times 10^{-7} \ , \\
\label{BRZ2}
BR_{\,\textrm{Exp}}(Z\to e^\pm \tau^\mp) & &< 9.8\times 10^{-6} \ , \\
\label{BRZ3}
BR_{\,\textrm{Exp}}(Z\to \mu^\pm \tau^\mp) && < 1.2\times 10^{-5} \ .
\end{eqnarray}
On the other hand, very stringent bounds arise also from leptonic three-body decays $l_A \to l_B \bar{l}_Cl_C$. The Particle Data Group reports the following limits:
\begin{eqnarray}
\label{BR3body1}
 BR_{\,\textrm{Exp}}(\mu \to e e^+e^-) && < 1.0\times 10^{-12} \ , \\
 \label{BR3body2}
 BR_{\,\textrm{Exp}}(\tau \to e e^+e^-) && < 2.7 \times 10^{-8}\ , \\
 \label{BR3body3}
 BR_{\,\textrm{Exp}}(\tau \to \mu e^+e^-) &&< 1.8 \times 10^{-8}\ , \\
 \label{BR3body4}
 BR_{\,\textrm{Exp}}(\tau \to \mu \mu^+\mu^-) && < 2.1 \times 10^{-8}\, .
\end{eqnarray}

Undoubtedly, the recent discovery of the Higgs boson at the Large Hadron Collider (LHC) opens up a new era in high-energy Higgs physics. In order to establish unambiguously that this scalar resonance corresponds to the Higgs boson predicted by the SM, many of the coming experiments in LHC will be focused on studying its decays into SM particles. The properties of this particle will also be under scrutiny at the International Linear Collider (ILC)~\cite{ILC}, which is currently at planning stage. This ambitious program of electron-positron collisions, operating with a center-of-mass energy in the range of $250-1000$ GeVs, will provide a clean environment to make studies beyond the capabilities of the LHC, expanding our knowledge of the SM by opening access to new physics that could eventually be unfolded through quantum fluctuations of SM observables. However, it is important to mention that in the case of Higgs LFV decays, their observation in the LHC is potentially as high as in the ILC. The reason is that these channels are relatively clean in this collider, especially the one involving the first two families. Besides that, Higgs bosons will be produced more copiously in the LHC than in the ILC. Effects of new heavy particles would show up more clearly in those processes that are forbidden or very suppressed in the SM. The LFV decays of the Higgs boson into pairs of distinct charged leptons, $H\to l_Al_B$, fall into the former type of cases. Although strictly forbidden in the SM, Higgs-mediated decays involving LFV arise naturally in the presence of extended Higgs sectors. This phenomenon has been the subject of important interest in the literature within the context of the two-Higgs doublet model~\cite{THDM}, supersymmetric theories~\cite{SUSY}, unparticle models~\cite{UPM}, 331 models~\cite{331}, seesaw models with Majorana massive neutrinos~\cite{SSM}, and in model-independent way using the effective Lagrangian technique~\cite{DT,EL}.

The purpose of this work is to study the LFV Higgs boson $H\to l^\pm_Al^\mp_B$ decays within the context of the Standard Model Extension (SME)~\cite{SME}, which is a SM extension that incorporates in a model-independent fashion both $CPT$ and Lorentz violation.  Although motivated from specific scenarios in the context of strings theory~\cite{KS}, general relativity with spontaneous symmetry breaking~\cite{QG} or field theories formulated in a noncommutative spacetime~\cite{Snyder,SWM,NCYM,NCSM}, this SME is beyond these specific ideas due to its generality, which is the main advantage of effective field theories. Thus this extension provides us with a powerful tool for investigating $CPT$ nonconservation and Lorentz violation in a model-independent way. In the aforementioned SME, terms of the form $T^{\mu_1,\ \cdots \mu_n}{\cal O}_{\mu_1,\ \cdots \mu_n}(x)$ are included. Each of these terms are required to be invariant under the SM gauge group, the Lorentz $ n- $tensors ${\cal O}_{\mu_1,\ \cdots \mu_n}(x)$ are assumed to be SM field dependent, and covariantly transform under both the particle and observer Lorentz transformations; however, the constants $T^{\mu_1,\ \cdots \mu_n}$ follow a covariant transformation under \emph{only} observer Lorentz transformations~\cite{SME, SME-NCSM}, hence the particle Lorentz group is broken. In this extended framework, the action can contain various $CPT$-odd terms, which necessarily implies Lorentz violation~\cite{CPTT}. Although the minimal version of the SME~\cite{SME} is constructed by adding to the SM Lagrangian new observer Lorentz invariant objects of the form described above, which are renormalizable in the Dyson's sense, it can be enlarged to include nonrenormalizable interactions~\cite{IT}. In this work, we will focus on a Yukawa sector extended by both renormalizable and nonrenormalizable Lorentz violating interactions, which directly induces the $Hl_Al_B$ vertex at tree level. We will show that Higgs-mediated LFV is quite suppressed in the context of the renormalizable version of the SME due to experimental constraints, but it can reach significant branching ratios in the context of its nonrenormalizable enlargement.

The rest of the paper has been organized as follows. In section~\ref{RSME} the phenomenon of LFV mediated by the photon, the $Z$ gauge boson, and the Higgs boson is studied in the context of the Yukawa sector of the renormalizable version of the SME. In this context, the experimental constraints on the photon-mediated transitions $l_A\to l_B\gamma$ are implemented to predict the LFV decays $H\to l_Al_B$ and  $Z\to l_Al_B$. Section~\ref{NRSME} is devoted to investigate the possible gauge and Lorentz observer invariant Yukawa-like operators of the lowest nonrenormalizable dimension that can generate Higgs--mediated LFV. In this enlarged version of the SME, the experimental constraints on the LFV three-body decays $l_A\to l_B\bar{l}_Cl_C$ are employed to predict the branching ratios of the $H\to l_Al_B$ and  $Z\to l_Al_B$ decays. Finally, in section~\ref{C} the conclusions are presented.
%-------------------------------------------------

\section{Lorentz violating Yukawa sector: renormalizable extension}
\label{RSME}
Throughout the paper we will be referring to the renormalizable and the nonrenormalizable versions of the SME. Let us clarify these concepts. Here the term renormalizable is used in the usual sense but with some caution. What we have called renormalizable version of the SME is given by a Lagrangian which comprises only interactions of canonical dimension less or equal than four, that is,
\begin{equation}
{\cal L}^{RV}_{SME}={\cal L}_{SM}+ \Delta {\cal L}^R\, ,
\end{equation}
where ${\cal L}_{SM}$ is the SM Lagrangian, while $\Delta {\cal L}^R$  contains $T^{\mu_1,\ \cdots \mu_n}{\cal O}_{\mu_1,\ \cdots \mu_n}(x)$, where ${\cal O}_{\mu_1,\ \cdots \mu_n}(x)$ represents $SU_C(3)\times SU_L(2)\times U_Y(1)$-invariant operators of dimension less or equal than four and $T^{\mu_1,\ \cdots \mu_n}$ are constant background fields. Although the divergent structure of this version has been explored at the one-loop level~\cite{QEDR,EWR,QCDR}, as far as we know its renormalizabilty at all orders has not been proved. As far as the nonrenormalizable version of the model is concerned, it incorporates all possible interactions consistent with observer Lorentz transformations and the standard gauge group. The corresponding effective Lagrangian can be written as follows:
\begin{equation}
{\cal L}^{NRV}_{SME}={\cal L}^{RV}_{SM}+ \Delta {\cal L}^{NR}\, ,
\end{equation}
where the $\Delta {\cal L}^{NR}$ term is given by a series, in principle infinite, which contains all allowed interactions of dimension higher than four.

In this section, we focus on the renormalizable extension to the leptonic Yukawa sector that induces LFV mediated by the Higgs boson. The only renormalizable extension of this sector is given by~\cite{SME}
\begin{equation}
{\cal L}^{CPT-even}_Y=-\frac{1}{2}Y'^{AB}_{\mu \nu}\bar{L}'_A \Phi \sigma^{\mu \nu}R'_B+\,\textrm{h.c.}\ ,
\end{equation}
where $\Phi$ is the $SU_L(2)$ Higgs doublet, whereas $L'$ and $R'$ are left--handed and right--handed lepton doublet and singlet of $SU_L(2)$, respectively. The dimensionless matrix $Y'$ is antisymmetric in the Lorentz indices but symmetric, although not necessarily Hermitian, in the flavor space. In the unitary gauge, the above Lagrangian can be written as follows:
\begin{equation}
{\cal L}^{CPT-even}_Y=-\frac{1}{2\sqrt{2}}\left(v+H\right)\bar{E}'_L Y'_{\mu \nu} \sigma^{\mu \nu}E'_R+\,\textrm{h.c.} \ ,
\end{equation}
where $E'=(e',\mu',\tau')$ is a vector in the flavor space. We now perform the change of basis from $(E'_L,E'_R)$ to the mass-eigenstate basis $(E_L,E_R)$ via the unitary transformation
\begin{eqnarray}
\label{Vtrans1}
E'_L && =V^l_LE_L \ , \\
\label{Vtrans2}
E'_R && =V^l_RE_R \ ,
\end{eqnarray}
which, as is well known, simultaneously diagonalizes the mass term and the Higgs-lepton interactions in the SM, but in this case introduces non-diagonal effects in the Lorentz violating extension,
\begin{equation}
{\cal L}^{CPT-even}_Y=-\frac{1}{2}\left(v+H\right)\bar{E}\left(Y_{\mu \nu}P_R+Y^\dag_{\mu \nu}P_L\right)\sigma^{\mu \nu} E\ ,
\end{equation}
where $Y_{\mu \nu}=V^{l\, \dag}_LY'_{\mu \nu}V^l_R$ and $P_{L,R}=(1\mp \gamma_5)/2$. Although the case of the most general $Y$ matrix may have interesting implications in some processes, such as the induction of  CP violation, for our purposes it is sufficient to assume it real and symmetric. This assumption considerably simplifies the analysis. Henceforth,
\begin{equation}
{\cal L}^{CPT-even}_Y=-\frac{1}{2}\left(v+H\right)\bar{E}Y_{\mu \nu}\sigma^{\mu \nu} E\ ,
\end{equation}
leads to a bilinear coupling $l_Al_B$, with vertex $-\frac{iv}{2}Y_{\mu \nu}\sigma^{\mu \nu}$, and contains a LFV coupling of the Higgs boson $Hl_Al_B$, whose vertex function is given by $-\frac{i}{2}Y_{\mu \nu}\sigma^{\mu \nu}$.

\subsection{The decay $H\to l^\pm _Al^\mp_B$}

\begin{figure}
\centering\includegraphics[width=4.5in]{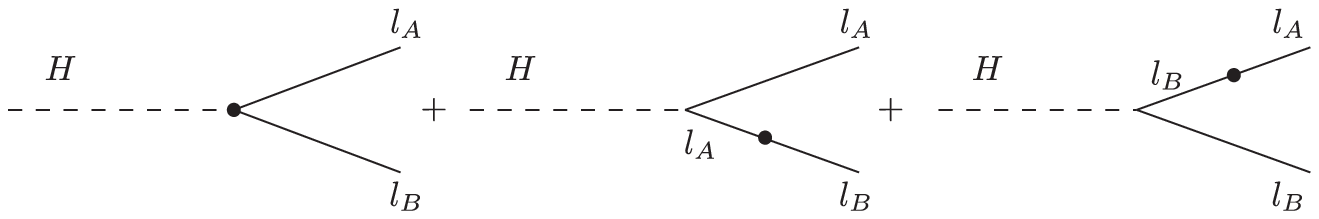}
\caption{\label{HDR}Diagrams contributing to the decay  $H\to l_Al_B$ in the context of a renormalizable extension of the Yukawa sector. Dots denote anomalous interactions.}
\end{figure}

The decay $H\to \bar{l}_Bl_A+\bar{l}_Al_B$ occurs via the Feynman diagrams shown in Fig.\ref{HDR}. Both the bilinear $l_Al_B$ and trilinear $Hl_Al_B$ couplings contribute to this process at tree level. The invariant amplitude for the decay $H\to \bar{l}_Bl_A$ can be written as follows:
\begin{eqnarray}
{\cal M}(H\to \bar{l}_Bl_A)&=&-\frac{i}{2}Y^{AB}_{\alpha \beta}\bar{v}(p_2,s_2)\Big[\sigma^{\alpha \beta}+\frac{m_B}{m^2_A-m^2_B}\left(\pFMSlash{p_1}+m_B\right)\sigma^{\alpha \beta}\nonumber \\
&&-\frac{m_A}{m^2_A-m^2_B}\sigma^{\alpha \beta}\left(\pFMSlash{p_2} +m_A\right)\Big]u(p_1,s_1)\ .
\end{eqnarray}
Notice that the contribution of the $Hl_Al_B$ vertex (first diagram in Fig.\ref{HDR}) is exactly cancelled by the contributions of the bilinear coupling $l_Al_B$. Once this amplitude is squared, the following branching ratio for the decay $H\to l^\pm_Al^\mp_B$  is obtained:
\begin{eqnarray}
\label{BR-HLL}
BR(H\to l^\pm_Al^\mp_B)&=&\frac{1}{4\pi}\left(\frac{m_H}{\Gamma_H}\right)\frac{m^4_H}{\left(m^2_A-m^2_B\right)^2}\nonumber \\
&\times &f(m_H,m_A,m_B)\big\|Y^{AB}_{\alpha \beta}Y^{AB}_{\lambda \rho}T^{\alpha \beta \lambda \rho}\big\|\ ,
\end{eqnarray}
where
\begin{equation}\fl
\qquad f(m_H,m_A,m_B)=\left[1-\left(\frac{m_A+m_B}{m_H}\right)^2\right]^{5/2}\left[1-\left(\frac{m_A-m_B}{m_H}\right)^2\right]^{1/2} \ ,
\end{equation}
and $ T^{\alpha \beta \lambda \rho} $ is given by
\begin{eqnarray}
T^{\alpha \beta \lambda \rho}  &=&\  \frac{m^2_B}{m^2_H-\left(m_A+m_B\right)^2}\left(g^{\alpha \lambda}g^{\beta \rho}-g^{\alpha \rho}g^{\beta \lambda}\right)\nonumber\\
 &&+\frac{m_B}{m_A}\left(A^{\alpha \beta \lambda \rho}+\frac{m^2_H-m^2_A-m^2_B}{m^2_H-\left(m_A+m_B\right)^2} B^{\alpha \beta \lambda \rho}\right) \nonumber \\
&&\ +\left(\frac{m_B}{m_A}\right)^2P^{\alpha \beta \lambda \rho}(p_1,p_1)+P^{\alpha \beta \lambda \rho}(p_2,p_2)\, .
\end{eqnarray}
Here, the coefficients $ A^{\alpha\beta\lambda\rho} $ and $ B^{\alpha\beta\lambda\rho} $ are
\begin{eqnarray}
A^{\alpha \beta \lambda \rho} &=& \frac{(p^\alpha_1p^\beta_2-p^\beta_1 p^\alpha_2)(p^\lambda_1p^\rho_2-p^\rho_1 p^\lambda_2)}{\left[m^2_H-\left(m_A+m_B\right)^2\right]^2} \ ,\\
B^{\alpha \beta \lambda \rho} &=& \frac{(p^\beta_1p^\lambda_2+p^\beta_2p^\lambda_1)g^{\alpha \rho}-( p^\alpha_1p^\lambda_2+p^\alpha_2p^\lambda_1)g^{\beta \rho}}{m^2_H-\left(m_A+m_B\right)^2}\nonumber \\
&&+\frac{( p^\alpha_1p^\rho_2+p^\alpha_2p^\rho_1) g^{\beta \lambda}-(p^\beta_1p^\rho_2+p^\beta_2p^\rho_1) g^{\alpha \lambda}}{m^2_H-\left(m_A+m_B\right)^2}\  ,
\end{eqnarray}
and $ P^{\alpha\beta\lambda\rho} $ explicitly reads as follows:
\begin{equation}\fl
\qquad P^{\alpha \beta \lambda\rho}\left(p_i,p_i\right)  =\frac{p^\beta_i p^\lambda_i g^{\alpha \rho}-p^\alpha_i p^\lambda_i g^{\beta \rho}+p^\alpha_i p^\rho_i g^{\beta \lambda}-p^\beta_i p^\rho_i g^{\alpha \lambda}}{m^2_H-\left(m_A+m_B\right)^2}\, , \, \, i=1,2 \ .
\end{equation}
Notice that all  these tensors are antisymmetric in the pairs of indices $\alpha \beta$ and $\lambda \rho$, and symmetric under the interchange $\alpha \beta \leftrightarrow \lambda \rho$, in agreement with the antisymmetry property of $ Y $ in its Lorentz indices and the structure of (\ref{BR-HLL}).

In the limit $m_B \to 0$ and depreciating $m_A$ against $m_H$ whenever possible, the branching ratio reduces to
\begin{eqnarray}
\label{BR-HLL-limit}
BR(H\to l^\pm_Al^\mp_B)&=&\frac{1}{4\pi}\left(\frac{m_H}{\Gamma_H}\right)\left(\frac{m_H}{m_A}\right)^4 \nonumber \\
&\times& \Big\|\left(\frac{2p_2}{m_H}\right)\cdot
\left(Y^{AB}gY^{AB}\right)\cdot \left(\frac{2p_2}{m_H}\right)\Big\|\ ,
\end{eqnarray}
where
\begin{equation}
p\cdot\left(Y^{AB}gY^{AB}\right)\cdot q \equiv p^\alpha Y^{AB}_{\alpha \beta}g^{\beta \lambda}Y^{AB \, \lambda \rho}q_\rho \ .
\end{equation}
In this kinematical limit, $E_1=E_2=|\textbf{p}_1|=|\textbf{p}_2|\equiv |\textbf{p}|=m_H/2$. Taking advantage of the property $Y^{AB}_{\alpha \beta}=-Y^{AB}_{\beta\alpha}$ we define the electric-like vector $Y^{AB}_{0i}=e^{AB}_i$, and the magnetic-like vector $Y^{AB}_{ij}=\epsilon_{ijk}b^{AB}_k$, just like the relation between the electric and magnetic fields with the corresponding electromagnetic strength tensor. Thus Eq. (\ref{BR-HLL-limit}) can be rewritten in terms of vectors \textbf{e}$^{AB}$ and \textbf{b}$^{AB}$ as follows:
\begin{eqnarray}
\Big \|f_1(\hat{\textbf{p}},\textbf{e}^{AB},\textbf{b}^{AB})\Big \| &\equiv&\Big\|\left(\frac{2p_2}{m_H}\right)\cdot
\left(Y^{AB}gY^{AB}\right)\cdot \left(\frac{2p_2}{m_H}\right)\Big\| \nonumber \\
&=&\Big\| (e^{AB})^2+(b^{AB})^2+2 \hat{\textbf{p}}\cdot (\textbf{e}^{AB}\times \textbf{b}^{AB})\nonumber \\
&&-(\hat{\textbf{p}}\cdot \textbf{e}^{AB})^2-(\hat{\textbf{p}}\cdot \textbf{b}^{AB})^2\Big\| \, ,
\end{eqnarray}
where $\hat{\textbf{p}}_2\equiv \frac{\textbf{p}_2}{|\textbf{p}_2|}$, $e^{AB}=|\textbf{e}^{AB}|$, and $b^{AB}=|\textbf{b}^{AB}|$. In terms of the electromagnetic-like vectors and the $\hat{\textbf{p}}_2$ direction, the Higgs branching ratio is
\begin{equation}
BR(H\to l^\pm_Al^\mp_B)=\frac{1}{4\pi}\left(\frac{m_H}{\Gamma_H}\right)\left(\frac{m_H}{m_A}\right)^4\Big \|f_1(\hat{\textbf{p}},\textbf{e}^{AB},\textbf{b}^{AB})\Big \| \ .
\end{equation}
In order to predict this branching ratio, we need to estimate the parameters in $Y^{AB}_{\alpha \beta}$. In the next subsection, we do this using experimental constraints.

\subsection{The decay $Z\to l^\pm_Al^\mp_B$}
\begin{figure}
\centering\includegraphics[width=2.5in]{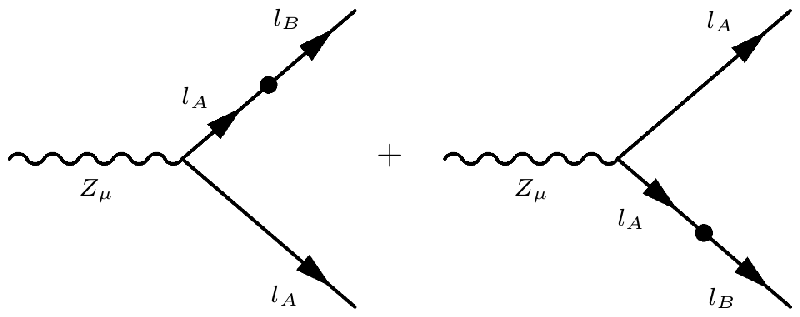}
\caption{\label{ZDR}Diagrams contributing to the decay $Z\to l_Al_B$ in the context of a renormalizable extension of the Yukawa sector. Dots denote anomalous interactions.}
\end{figure}

As the branching ratios (\ref{BRgm1}-\ref{BR3body4}) indicate, the current bounds on the decays $Z\to \bar{l}_Bl_A+\bar{l}_Al_B$ are not as severe as those imposed on the electromagnetic transitions, or as those set on the three-body decays of charged leptons. So, there may still be a window to observe LFV mediated by the $Z$ gauge boson. In this subsection, we study the impact of the renormalizable extension of the Yukawa sector on these $Z$ decays. These decays have contributions at tree level due to the bilinear couplings $l_Al_B$, as it is shown in Fig.~\ref{ZDR}. Following these diagrams, the invariant amplitude for the decay $Z\to \bar{l}_Bl_A$ is
\begin{equation}
{\cal M}(Z\to \bar{l}_Bl_A)=\frac{i}{2}\frac{m_Z}{m^2_A-m^2_B}\bar{v}(p_2,s_2)\Lambda_\mu u(p_1,s_1) \epsilon^\mu(p,\lambda)\ ,
\end{equation}
where
\begin{eqnarray}
\Lambda_\mu&=&Y^{AB}_{\alpha \beta}\Big[\gamma_\mu\left(g^l_V-g^l_A\gamma_5\right)\left(\pFMSlash{p_1}+m_B\right)\sigma^{\alpha \beta}\nonumber \\
&&-\sigma^{\alpha \beta}\left(\pFMSlash{p_2}+m_A\right)\gamma_\mu\left(g^l_V-g^l_A\gamma_5\right) \Big]\ ,
\end{eqnarray}
moreover $g^l_V=\frac{1}{2}+2s^2_W$, with $s_W$ standing for the sine of the weak angle, and $g^l_A=\frac{1}{2}$. Squaring this amplitude leads to,
\begin{equation}
|\bar{{\cal M}}|^2=\left(\frac{m_Z}{m_A}\right)^4 F(m_Z,m_A,m_B)\ ,
\end{equation}
with $ F(m_Z,m_A,m_B)$ a rather cumbersome expression whose explicit form is not necessary for our purposes. Working in the limits  $\frac{m_A}{m_Z} \rightarrow 0$ and $\frac{m_B}{m_Z} \rightarrow 0$, which imply the kinematical conditions $E_1=E_2=|\textbf{p}_1|=|\textbf{p}_2|\equiv |\textbf{p}|=m_Z/2$, the branching ratio for the decay $Z\to l^\pm_Al^\mp_B$ becomes
\begin{eqnarray}
BR(Z\to l^\pm_Al^\mp_B)&=&\left(\frac{(g^l_V)^2+(g^l_A)^2}{\pi}\right)\left(\frac{m_Z}{m_A}\right)^4\left(\frac{m_Z}{\Gamma_Z}\right)\nonumber \\
&\times &\Big \|Y^{AB}_{\alpha \beta}Y^{AB}_{\lambda \rho}{\cal Z}^{\alpha \beta \lambda \rho}\Big \|\ ,
\end{eqnarray}
where the tensor ${\cal Z}^{\alpha \beta \, \lambda \rho}$ satisfies ${\cal Z}^{\alpha \beta \, \lambda \rho}=-{\cal Z}^{\beta \alpha \, \lambda \rho}=-{\cal Z}^{\alpha \beta \, \rho \lambda }=+{\cal Z}^{\lambda \rho \, \alpha \beta}$ and is given by
\begin{eqnarray}
{\cal Z}^{\alpha \beta \lambda \rho}&=&\frac{1}{m^2_Z}\Big[(p^\beta_1p^\rho_1+p^\beta_2 p^\rho_2)g^{\alpha \lambda}-(p^\alpha_1p^\rho_1+p^\alpha_2 p^\rho_2)g^{\beta \lambda}\nonumber \\
&&+ (p^\alpha_1p^\lambda_1+p^\alpha_2 p^\lambda_2)g^{\beta \rho}-(p^\beta_1p^\lambda_1+p^\beta_2 p^\lambda_2)g^{\alpha \rho}\Big]\ .
\end{eqnarray}
The index symmetry properties of ${\cal Z}^{\alpha \beta \, \lambda \rho}  $ and the already introduced kinematical approximation yield a branching ratio for the $Z$ decay in terms of the function $f_1(\hat{\textbf{p}},\textbf{e}^{AB},\textbf{b}^{AB})$, which is also involved in the branching ratio Eq.~(\ref{BR-HLL-limit}) of the Higgs decay. Explicitly,
\begin{eqnarray}
\label{BR-Z-limit}
BR(Z\to l^\pm_Al^\mp_B)&=&2\left(\frac{(g^l_V)^2+(g^l_A)^2}{\pi}\right)\left(\frac{m_Z}{m_A}\right)^4\left(\frac{m_Z}{\Gamma_Z}\right)\nonumber \\
&\times & \Big \|f_1(\hat{\textbf{p}},\textbf{e}^{AB},\textbf{b}^{AB})\Big \| \ .
\end{eqnarray}

\subsection{The electromagnetic decay $l_A \to l_B\gamma$}
\begin{figure}
\centering\includegraphics[scale=.85]{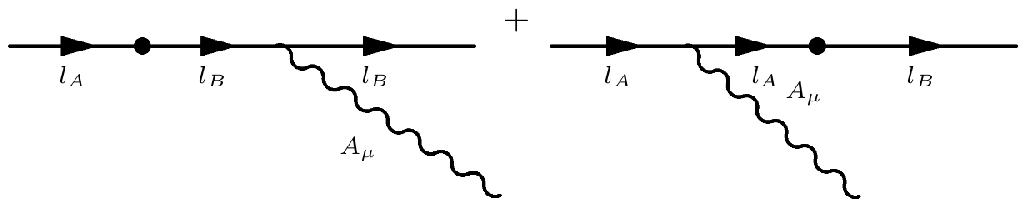}
\caption{\label{EDR}Diagrams contributing to the $l_A\to l_B\gamma$ decay in the context of a renormalizable extension of the Yukawa sector. Dots denote anomalous interactions.}
\end{figure}

The most stringent bounds on LFV come from the electromagnetic decays $l_A \to l_B\gamma$ and the three-body charged lepton decays $l_A\to l_B \bar{l}_Cl_C$ (see (\ref{BRgm1}-\ref{BRgm3}) and (\ref{BR3body1}-\ref{BR3body4})), both being of the same order of magnitude. Since the two-body decays $l_A \to l_B\gamma$ under consideration are induced at tree level, due to the presence of the bilinear terms $l_Al_B$ (see Fig.~\ref{EDR}), it is clear that this type of decay dominates on the corresponding three-body one. Accordingly, we proceed to calculate the branching ratio for the decays $l_A \to l_B\gamma$ and use the experimental limits on them to constraint the function $f_1(\hat{\textbf{p}},\textbf{e}^{AB},\textbf{b}^{AB})$ that consistently appears in the branching ratios~(\ref{BR-HLL-limit}) and (\ref{BR-Z-limit}).

The invariant amplitude for the decay $l_A \to l_B\gamma$  is given by
\begin{equation}
{\cal M}(l_A \to l_B \gamma)=\frac{is_W m_W}{m^2_A-m^2_B}\, \bar{u}(p_2,s_2)\Gamma_\mu u(p_1,s_1)\, \epsilon^{\mu *}(q,\lambda)\ ,
\end{equation}
where
\begin{equation}
\Gamma_\mu=Y^{AB}_{\alpha \beta}\left[\sigma^{\alpha \beta}(\pFMSlash{p_2}+m_A)\gamma_\mu -\gamma_\mu(\pFMSlash{p_1}+m_B)\sigma^{\alpha \beta} \right] \ .
\end{equation}
Although it is not evident, it can be shown that this amplitude preserves gauge invariance as it satisfies the Ward identity
\begin{equation}
q^\mu \Gamma_\mu=0\ .
\end{equation}

Once the amplitude is squared, the corresponding branching ratio can be written as follows:
\begin{eqnarray}
BR(l_A\to l_B\gamma)&=&\left(\frac{s^2_W}{2\pi}\right)\left(\frac{m_W}{m_A}\right)^2\left(\frac{m_A}{\Gamma_A}\right)\left(1-\frac{m^2_B}{m^2_A}\right)\nonumber \\
&\times &\Big\| Y^{AB}_{\alpha \beta}Y^{AB}_{\lambda \rho}R^{\alpha \beta \lambda \rho}\Big \| \ ,
\end{eqnarray}
where $\Gamma_A$ is the total decay width of the charged lepton $l_A$ and
\begin{equation}\fl
\qquad R^{\alpha \beta \lambda \rho}=g^{\alpha \lambda}g^{\beta \rho}-g^{\alpha \rho}g^{\beta \lambda}+\frac{m^4_A}{\left(m^2_A-m^2_B\right)^2}A^{\alpha \beta \lambda \rho}+\frac{m^4_B}{\left(m^2_A-m^2_B\right)^2}B^{\alpha \beta \lambda \rho}\ ,
\end{equation}
where
\begin{eqnarray}
A^{\alpha \beta \lambda \rho} & =&\frac{{\cal A}^{\alpha \beta \lambda \rho}(p_1,p_2)}{m^2_A} \ , \\
B^{\alpha \beta \lambda \rho} & =&\frac{{\cal A}^{\alpha \beta \lambda \rho}(p_2,p_1)}{m^2_B} \ ,
\end{eqnarray}
here
\begin{eqnarray}
{\cal A}^{\alpha \beta \lambda \rho}(p_{i},p_{j})  &= & \left[p^\beta_j \left(p_i-3p_j\right)^\lambda+ p^\beta_i\left(p_i+p_j\right)^\lambda \right]g^{\alpha \rho}\nonumber \\
&&- \left[p^\alpha_j \left(p_i-3p_j\right)^\lambda+ p^\alpha_i\left(p_i+p_j\right)^\lambda \right]g^{\beta \rho} \nonumber \\
&&-\left[p^\beta_j \left(p_i-3p_j\right)^\rho+ p^\beta_i\left(p_i+p_j\right)^\rho \right]g^{\alpha \lambda}\nonumber \\
&&+\left[p^\alpha_j \left(p_i-3p_j\right)^\rho + p^\alpha_i\left(p_i+p_j\right)^\rho \right]g^{\beta \lambda}\ .
\end{eqnarray}
In this expression, $i,j=1,2$. Notice that ${\cal A}^{\alpha \beta \, \lambda \rho}=-{\cal A}^{\beta \alpha \, \lambda \rho}=-{\cal A}^{\alpha \beta \, \rho \lambda}=+{\cal A}^{\beta \alpha \, \rho \lambda}$.

Depreciating $m_B$ against $m_A$, the branching ratio acquires a simpler form, namely
\begin{eqnarray}
\label{BR-gm-limit}
BR(l_A\to l_B\gamma)&=&\left(\frac{s^2_W}{2\pi}\right)\left(\frac{m_W}{m_A}\right)^2\left(\frac{m_A}{\Gamma_A}\right)\nonumber \\
&\times &\Big\| Y^{AB}_{\alpha \beta}Y^{AB}_{\lambda \rho}\left(g^{\alpha \lambda}g^{\beta \rho}-g^{\alpha \rho}g^{\beta \lambda}+A^{\alpha \beta \lambda \rho}\right)\Big \| \ .
\end{eqnarray}
Just as for the decays  $H\to l^\mp_A l^\pm_B$ and $Z\to l^\mp_A l^\pm_B$, we can write the branching ratio Eq.~(\ref{BR-gm-limit}) in terms of the electric- and magnetic-like vectors \textbf{e}$^{AB}$ and \textbf{b}$ ^{AB} $ contained in $Y^{AB}_{\alpha \beta}$ as
\begin{equation}
BR(l_A\to l_B\gamma)=\left(\frac{s^2_W}{2\pi}\right)\left(\frac{m_W}{m_A}\right)^2\left(\frac{m_A}{\Gamma_A}\right)
\Big\|f_2(\hat{\textbf{p}},\textbf{e}^{AB},\textbf{b}^{AB})\Big \|\ ,
\end{equation}
where
\begin{eqnarray}
\Big\|f_2(\hat{\textbf{p}},\textbf{e}^{AB},\textbf{b}^{AB})\Big \|&=&\Big\| (e^{AB})^2+(b^{AB})^2-2 \hat{\textbf{p}}\cdot (\textbf{e}^{AB}\times \textbf{b}^{AB})\nonumber \\
&&+3(\hat{\textbf{p}}\cdot \textbf{e}^{AB})^2+3(\hat{\textbf{p}}\cdot \textbf{b}^{AB})^2\Big\| \ ,
\end{eqnarray}
and $\hat{\textbf{p}}$ stands for the normalized momentum vector of the lepton $l_B$.

%---------------------------------------------------

\subsection{Discussion}
In order to bound the branching ratios for the  decays $H\to l^\mp_A l^\pm_B$ and $Z\to l^\mp_A l^\pm_B$, we will use the experimental limits on the electromagnetic transitions  $l_A\to l_B \gamma$ given in the introduction. So, we demand
\begin{equation}
BR(l_A\to l_B\gamma)<BR_{\,\textrm{Exp}}(l_A\to l_B\gamma)\ ,
\end{equation}
which leads to
\begin{equation}\fl
\qquad \Big\|f_2(\hat{\textbf{p}},\textbf{e}^{AB},\textbf{b}^{AB})\Big \|<\left(\frac{2\pi}{s^2_W}\right)\left(\frac{m_A}{m_W}\right)^2\left(\frac{\Gamma_A}{m_A}\right)BR_{\,\textrm{Exp}}(l_A\to l_B\gamma)\ .
\end{equation}
Using the experimental limits (\ref{BRgm1}-\ref{BRgm3}), one obtains the following severe bounds:
\begin{eqnarray}
\label{f2bounds}
\Big\|f_2(\hat{\textbf{p}},\textbf{e}^{\mu e},\textbf{b}^{\mu e})\Big \| && <7.1\times 10^{-35} \ ,\\
\Big\|f_2(\hat{\textbf{p}},\textbf{e}^{\tau \mu},\textbf{b}^{\tau \mu})\Big \| && < 1.0\times 10^{-21} \ .
\end{eqnarray}

On the other hand, to predict the branching ratios $BR(H\to l^\pm_{A}l^\mp_B)$ and $BR(Z\to l^\pm_{A}l^\mp_B)$, we need to estimate the value of $\Big\|f_1(\hat{\textbf{p}},\textbf{e}^{AB},\textbf{b}^{AB})\Big \|$ relative to $\Big\|f_2(\hat{\textbf{p}},\textbf{e}^{AB},\textbf{b}^{AB})\Big \|$. To this end, we consider the following scenarios:
\begin{itemize}
\item Scenario 1: $\textbf{e}^{AB}\times \textbf{b}^{AB}\neq 0$. Within this scenario, we can consider the following two possibilities

\noindent (a) The unitary vector $\hat{\textbf{p}}$ lies on the plane defined by $\textbf{e}^{AB}$ and $\textbf{b}^{AB}$, $\hat{\textbf{p}}\cdot(\textbf{e}^{AB}\times \textbf{b}^{AB})=0$. So $\| f_{1}( \hat{\textbf{p}}, \textbf{e}^{AB}, \textbf{b}^{AB}) \|$ and $\| f_{2}( \hat{\textbf{p}}, \textbf{e}^{AB}, \textbf{b}^{AB})  \|$  reduced to

\begin{eqnarray*}
\!\!\!\!\!\!\!\!\!\!\!\!\!\!\!\!\!\!\!\!\Big \| f_{1}( \hat{\textbf{p}}, \textbf{e}^{AB}, \textbf{b}^{AB}) \Big \|&=& \Big \| (e^{AB})^{2}+(b^{AB})^{2}-(\hat{\textbf{p}}\cdot\textbf{e}^{AB})^{2} -(\hat{\textbf{p}}\cdot\textbf{b}^{AB})^{2} \Big \| \ , \\
\!\!\!\!\!\!\!\!\!\!\!\!\!\!\!\!\!\!\!\!\Big \| f_{2}( \hat{\textbf{p}}, \textbf{e}^{AB}, \textbf{b}^{AB})  \Big \|&=& \Big \| (e^{AB})^{2}+(b^{AB})^{2}+3(\hat{\textbf{p}}\cdot\textbf{e}^{AB})^{2} +3(\hat{\textbf{p}}\cdot\textbf{b}^{AB})^{2} \|\ ,
\end{eqnarray*}	
thus
\begin{equation}
\label{b}
\Big \| f_{2}( \hat{\textbf{p}}, \textbf{e}^{AB}, \textbf{b}^{AB})  \Big \|>\Big \| f_{1}( \hat{\textbf{p}}, \textbf{e}^{AB}, \textbf{b}^{AB}) \Big \|\ .
\end{equation}

\noindent (b) The unitary vector $\hat{\textbf{p}}$ is anti-parallel to $(\textbf{e}^{AB}\times\textbf{b}^{AB})$, then $\hat{\textbf{p}}\cdot\textbf{e}^{AB}=\hat{\textbf{p}}\cdot\textbf{b}^{AB}=0$. And we have the following reductions:
	\begin{eqnarray*}
\Big\| f_{1}(\hat{\textbf{p}},\textbf{e}^{AB}, \textbf{b}^{AB}) \Big \| =\Big \| (e^{AB})^{2}+(b^{AB})^{2}-2\vert\hat{\textbf{p}}\vert\,\vert\textbf{e}^{AB}\times \textbf{b}^{AB}\vert \Big \| \,,\\
\Big \| f_{2}(\hat{\textbf{p}},\textbf{e}^{AB}, \textbf{b}^{AB})\Big \| =\Big \| (e^{AB})^{2}+(b^{AB})^{2}+2\vert\hat{\textbf{p}}\vert\,\vert\textbf{e}^{AB}\times \textbf{b}^{AB}\vert \Big \| \,,
	\end{eqnarray*}
so we have again the relation (\ref{b}).

\item Scenario 2: $\textbf{e}^{AB}\times \textbf{b}^{AB}= 0$. In this case, $\| f_{1}(\hat{\textbf{p}}, \textbf{e}^{AB}, \textbf{b}^{AB})  \|$ and $\| f_{2}(\hat{\textbf{p}}, \textbf{e}^{AB}, \textbf{b}^{AB})  \|$ are reduced to
\begin{eqnarray*}
\!\!\!\!\!\!\!\!\!\!\!\!\!\!\!\!\!\!\!\!\Big \| f_{1}(\hat{\textbf{p}}, \textbf{e}^{AB}, \textbf{b}^{AB})  \Big \| &=& \Big \| (e^{AB})^{2}+(b^{AB})^{2}-(\hat{\textbf{p}}\cdot\textbf{e}^{AB})^{2} -(\hat{\textbf{p}}\cdot\textbf{b}^{AB})^{2} \Big \| \,, \\
\!\!\!\!\!\!\!\!\!\!\!\!\!\!\!\!\!\!\!\!\Big \| f_{2}(\hat{\textbf{p}}, \textbf{e}^{AB}, \textbf{b}^{AB})  \Big \| &=& \Big \| (e^{AB})^{2}+(b^{AB})^{2}+3(\hat{\textbf{p}}\cdot\textbf{e}^{AB})^{2} +3(\hat{\textbf{p}}\cdot\textbf{b}^{AB})^{2} \Big \|\ ,
\end{eqnarray*}	
and again we got the result given by Eq.(\ref{b}).
\end{itemize}

From the above results, we conclude that whenever any of the previous scenarios are met, we have
\begin{eqnarray}
BR(H\to l^\pm_{A}l^\mp_B)&<&\frac{1}{4\pi}\left(\frac{m_H}{\Gamma_H}\right)\left(\frac{m_H}{m_A}\right)^4\Big \|f_2(\hat{\textbf{p}},\textbf{e}^{AB},\textbf{b}^{AB})\Big \| \, , \\
BR(Z\to l^\pm_Al^\mp_B)&<& 2\left(\frac{(g^l_V)^2+(g^l_A)^2}{\pi}\right)\left(\frac{m_Z}{\Gamma_Z}\right)\left(\frac{m_Z}{m_A}\right)^4\nonumber \\
&&\times  \Big \|f_2(\hat{\textbf{p}},\textbf{e}^{AB},\textbf{b}^{AB})\Big \| \, .
\end{eqnarray}
Comparing these inequalities with the general bounds (\ref{f2bounds}) on $ \| f_{2}( \hat{\textbf{p}}, \textbf{e}^{AB}, \textbf{b}^{AB})  \|$, imply the following Higgs branching ratios:
\begin{eqnarray}
BR(H\to \mu^\pm e^\mp) && <4.3\times 10^{-19}\ , \label{BRHiggs-bounds1}\\
BR(H\to \tau^\pm \mu^\mp)&&< 4.2\times 10^{-11}\ .\label{BRHiggs-bounds2}
\end{eqnarray}
In a similar fashion,  the corresponding branching ratios of the $Z$ gauge boson are given by
\begin{eqnarray}
BR(Z\to \mu^\pm e^\mp)&&< 2.5\times 10^{-21}\ , \label{BRZ-bounds1}\\
BR(Z\to \tau^\pm \mu^\mp)&&< 2.0\times 10^{-12}\ . \label{BRZ-bounds2}
\end{eqnarray}
From these results, it is evident that signals of lepton flavor violation mediated by the Higgs or the $Z$ gauge boson are severely suppressed if induced in a renormalizable context of Lorentz violation. Behind this behavior lies the fact that both the Higgs and $Z$ boson decays are induced at the same order that the electromagnetic ones, which are severely constrained by the experiment. Below, we will show that a quite different scenario emerges  when the SME is enlarged to comprise nonrenormalizable lepton flavor violating effects.

\section{Lorentz violating Yukawa sector: nonrenormalizable extension}
\label{NRSME}
In this section, we explore the implications of a Lorentz violating extension to the Yukawa sector beyond the renormalizable structure. The idea is to introduce an extension of the Yukawa sector via the lowest possible dimension terms that generate LFV mediated by the Higgs boson (and also perhaps by the $Z$ gauge boson) at tree level, but not by the photon. Notice that this effect cannot be generated by the photon at tree level, but only at one-loop or higher orders. In other words, we will consider a Lorenz violating nonrenormalizable extension of the Yukawa sector where the LFV electromagnetic transitions are naturally suppressed in a perturbative context. It is not difficult to convince oneself that the
unique\footnote{There is also the dimension-five $SU_L(2)\times U_Y(1)$-invariant operator $\overline{D^\alpha L'}_A\Phi l'_{RB}+\bar{L'}_A\Phi \overline{D^\alpha l'}_{RB}$; however, this is related to Eq.~(\ref{LNRY1}) via a surface term.} lowest dimension extension of the Yukawa sector, with the properties specified above, is given by the following dimension-five $SU_L(2)\times U_Y(1)$--invariant Lagrangian
\begin{equation}\label{LNRY1}
{\cal L}^{NR}_Y=-Y'^{AB}_{\alpha}\bar{L'}_AD^\alpha \Phi l'_{RB}+\,\textrm{h.c.} \ ,
\end{equation}
where $D^\alpha$ is the covariant derivative of the electroweak group in the doublet representation. In this expression, $Y'^{AB}_\alpha$ is a matrix in the flavor space with units of inverse of mass. In the unitary gauge and after using the standard unitary mapping that transforms gauge fields into mass eigenstate fields (see Eqs.~(\ref{Vtrans1},\ref{Vtrans2})), one obtains
\begin{equation}\label{LNRY}
{\cal L}^{NR}_Y=-\partial^\alpha H \, \bar{E}Y_\alpha E-im_Zc_{2W}\left(1+\frac{H}{v}\right)Z^\alpha \, \bar{E}Y_\alpha \gamma_5 E \ ,
\end{equation}
where $Y_\alpha=\frac{1}{\sqrt{2}}V^{l\dag}_L Y'_\alpha V^l_R$ is the matrix responsible for LFV. As in the renormalizable case, it will be assumed that the matrix $ Y_{\alpha} $ is real and symmetric. In addition, we introduced the shorthand notation $c_{2W}\equiv c^2_W-s^2_W$, with $c_W$ stands for cosine of the weak angle.

Directly from the structure of $ {\cal L}^{NR}_{Y} $, the LFV electromagnetic transitions are naturally suppressed as they first arise at one-loop level, whereas LFV effects mediated by the Higgs boson or the $Z$ gauge boson are induced at tree level. The respective vertex functions for the $H (p)l_Al_B$ and $Zl_Al_B$  couplings are $ip^\alpha Y^{AB}_\alpha$ and $m_Zc_{2W}\gamma_5 Y^{AB}_\alpha$. Hence we are now in position to calculate branching ratios of the different decays we are interested in.

\subsection{The decay $H\to l^\pm_Al^\mp_B$}
The amplitude for the decay $H\to \bar{l}_Bl_A$ is given by
\begin{eqnarray}
\label{MHiggs}
{\cal M}(H\to \bar{l}_Bl_A)&=&i\left(p^\alpha Y^{AB}_\alpha\right)\, \left[\bar{v}(p_2,s_2)u(p_1,s_1)\right]\nonumber \\
 &=&i\left(p^0 Y^{AB}_0\right)\, \left[\bar{v}(p_2,s_2)u(p_1,s_1)\right]\nonumber \\
 &=&i\left(m_HY^{AB}_0\right)\, \left[\bar{v}(p_2,s_2)u(p_1,s_1)\right]\, ,
\end{eqnarray}
where the last two expressions are valid at the rest frame of the Higgs boson. The Lagrangian that defines the Lorentz violating and
nonrenormalizable extension of the Yukawa sector, Eq.~(\ref{LNRY}), can be thought as an effective Lagrangian derived from a
fundamental Lorentz-invariant theory that contains interactions among a certain well behaved vector field $ Y_{\alpha}(x) $, the leptonic
fields and the Higgs boson, where the breakdown of the Lorentz symmetry arises from a timelike nonzero expectation value, $\langle
Y_{\alpha} \rangle_{0}\equiv (Y^{AB}_{0},0,0,0)$, acquired by $ Y_{\alpha}(x) $ at the  rest frame of the Higgs boson\footnote{Similar
ideas have been used in the so-called bumblebee models, which consider that the photons and gravitons could emerge as Goldstone bosons
from a spontaneous symmetry breaking that incorporates vacuum expectation values of the corresponding gauge fields~\cite{SSBLS}.}. As
neither such a fundamental theory or symmetry breaking mechanism is relevant for our discussion, we will assume the constant time-like
vector  $Y^{AB}_{\alpha}=(Y^{AB}_{0},0,0,0)$ as an effective coupling constant.

In order to consider the complete process  $H \to \bar{l}_Bl_A+\bar{l}_Al_B$ a factor of 2 must be included in Eq.~(\ref{MHiggs}) after the amplitude is squared; therefore, the branching ratio for this decay can be written as follows:
\begin{equation}
BR(H \to l^\pm_Al^\mp_B)=\frac{1}{4\pi} \left(\frac{m_H}{\Gamma_H}\right)f(m_H,m_A,m_B)\left(m_HY^{AB}_0\right)^2 \ ,
\end{equation}
where
\begin{eqnarray}
f(m_H,m_A,m_B)&=&\left(1-\frac{m^2_A+m^2_B}{m^2_H}\right)\nonumber \\
&\times&\sqrt{1-\left(\frac{m_A+m_B}{m_H}\right)^2}\sqrt{1-\left(\frac{m_A-m_B}{m_H}\right)^2}\ .
\end{eqnarray}
In the limit where the lepton masses $ m_{A} $ and $ m_{B} $ are disregarded against the Higgs mass $ m_{H} $, the function $f(m_H,m_A,m_B)\to 1$ so the branching ratio acquires the following simple form:
\begin{equation}
BR(H \to l^\pm _Al^\mp_B)=\frac{1}{4\pi} \left(\frac{m_H}{\Gamma_H}\right)\left(m_HY^{AB}_0\right)^2 \ .
\end{equation}

\subsection{The decay $Z\to l^\pm _Al^\mp_B$}
In a similar fashion, an exact calculation of the branching ratio for the decay $Z\to \bar{l}_Bl_A+\bar{l}_Al_B$, leads to
\begin{eqnarray}
BR(Z\to \bar{l}_Bl_A+\bar{l}_Al_B)&=&\frac{c^2_{2W}}{12\pi}\left(\frac{m_Z}{\Gamma_Z}\right)g(m_Z,m_A,m_B)\nonumber \\
&\times &\Big \|m^2_Z Y^{AB}_{\alpha }Y^{AB\, \alpha}+\left(p^\alpha Y^{AB}_\alpha\right)^2\Big \|\ ,
\end{eqnarray}
where $p^\alpha=(p^0,0,0,0)$ is the momentum of the $Z$ gauge boson in its rest frame, and the function $g(m_Z,m_A,m_B)$ is given by
\begin{equation}\fl
\qquad g(m_Z,m_A,m_B)=\left[1-\left(\frac{m_A-m_B}{m_Z}\right)^2\right]^{3/2}\left[1-\left(\frac{m_A+m_B}{m_Z}\right)^2\right]^{1/2}\ .
\end{equation}
In the limit $(m_A/m_Z), (m_B/m_Z) \to 0$, and taking $Y^{AB}_\alpha=(Y^{AB}_0,0,0,0)$, the branching ratio reduces to
\begin{equation}
BR(Z\to l^\pm_Al^\mp_B)=\frac{c^2_{2W}}{6\pi}\left(\frac{m_Z}{\Gamma_Z}\right)\left(m_ZY^{AB}_0\right)^2 \ .
\end{equation}

\subsection{The $l_A \to l_B \bar{l}_C l_C$ decay}
\begin{figure}
\centering\includegraphics[scale=.85]{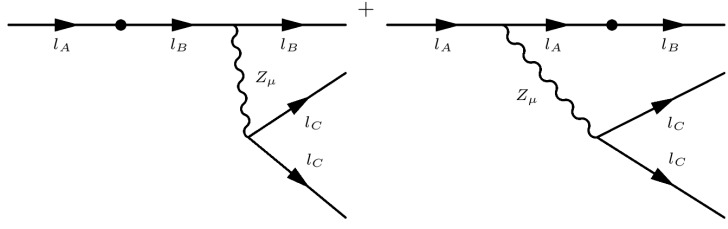}
\caption{\label{TBDNR}Diagrams contributing to the $l_A \to l_B \bar{l}_C l_C$ decay in the context of a nonrenormalizable extension of the Yukawa sector. Dots denote anomalous interactions.}
\end{figure}

In this \emph{non}renormalizable extension of the Yukawa sector, electromagnetic transitions that involve LFV arise at one-loop level in contrast to the renormalizable case where they arise at tree level. Nevertheless, from the extension Eq.~(\ref{LNRY}), we can calculate the
three-body decay $l_A \to l_B \bar{l}_C l_C$ induced at tree-level by the $Z$ gauge boson, this decay leads to the most important bound on
the $Y^{AB}_0$ scale\footnote{This decay can also be induced by the Higgs boson, but due to the small SM coupling in $H\bar{l}_Cl_C$,
this contribution is quite suppressed in comparison to the $Z$ gauge boson contribution.}. In accordance with the diagrams in the Fig.
\ref{TBDNR}, the invariant amplitude for the decay $l_A \to l_B \bar{l}_C l_C$ is
\begin{eqnarray}
{\cal M}(l_A \to l_B \bar{l}_C l_C)&=&-\frac{gc_{2W}Y^{AB}_\alpha}{2c_Wm_Z}\left[\bar{u}(p_2,s_2)\gamma_5 u(p_1,s_1)\right]\nonumber \\
&\times &\left[\bar{v}(q_2,s_2)\gamma^\alpha (g^l_V-g^l_A\gamma_5)u(q_1,s_1)\right]\ ,
\end{eqnarray}
where we have neglected the contribution of the longitudinal component of the $Z$ propagator, as it is proportional to $(m_A/m_Z)^2$. In the limit $m_B,m_C \to 0$, the branching ratio for this decay can be rewritten as follows:
\begin{equation}\label{BR3body-integral}
BR(l_A \to l_B \bar{l}_C l_C)=\frac{1}{256\pi^3} \left(\frac{m_A}{\Gamma_A}\right)\int^1_0 dx \int^{1-x}_0 dy |\bar{{\cal M}}|^2 \ ,
\end{equation}
where
\begin{equation}
|\bar{{\cal M}}|^2=\frac{g^2c^2_{2W}}{2c^2_W}\left[(g^l_V)^2+(g^l_A)^2\right]\left(\frac{m_A}{m_Z}\right)^4\Big \| m_Z Y^{AB}_0\Big \|^2 f(x,y)\ ,
\end{equation}
and
\begin{equation}
f(x,y)=(2-x-y)(1-x-y+xy) \ .
\end{equation}
Solving the integral in the Eq.~(\ref{BR3body-integral}),
\begin{eqnarray}\fl
\label{TBD}
\qquad BR(l_A \to l_B \bar{l}_C l_C)&=&\left(\frac{3}{10}\right)\left(\frac{\alpha}{64\pi^2}\right)\left(\frac{c_{2W}}{s_{2W}}\right)^2\nonumber \\
&\times &\left[(g^l_V)^2+(g^l_A)^2\right] \left(\frac{m_A}{\Gamma_A}\right)\left(\frac{m_A}{m_Z}\right)^4\Big \| m_Z Y^{AB}_0\Big \|^2\ .
\end{eqnarray}

\subsection{Discussion}
We now use the experimental constraints on the three-body decays of charged leptons to get bounds on the $Y^{AB}_0$ scale. By demanding
\begin{equation}
BR(l_A \to l_B \bar{l}_C l_C)<BR_{\,\textrm{Exp}}(l_A \to l_B \bar{l}_C l_C)\ ,
\end{equation}
one obtains
\begin{eqnarray}
\Big \| m_Z Y^{AB}_0\Big \|^2&<&\left(\frac{10}{3}\right)\left(\frac{64\pi^2}{\alpha}\right)\left(\frac{s_{2W}}{c_{2W}}\right)^2 \frac{1}{(g^l_V)^2+(g^l_A)^2}\nonumber \\
 &\times &\left(\frac{\Gamma_A}{m_A}\right)\left(\frac{m_Z}{m_A}\right)^4BR_{\,\textrm{Exp}}(l_A \to l_B \bar{l}_C l_C)\ .
\end{eqnarray}
So, using the experimental limits~\cite{PDG} on these decays, the following bounds are obtained:
\begin{eqnarray}
\label{B1}
\Big \| m_Z Y^{\mu e}_0\Big \|^2&<& 1.2\times 10^{-12} \, ,\\
\label{B2}
\Big \| m_Z Y^{\tau \mu}_0\Big \|^2&<& 2.0\times 10^{-7} \, .
\end{eqnarray}
These bounds allow us to estimate the branching ratios of the LFV decays of the Higgs and $Z$ bosons.

Considering the Higgs decay, we obtain the following limits:
\begin{eqnarray}
BR(H\to \mu^\pm e^\mp)&<&4.1\times 10^{-9} \, , \\
BR(H\to \tau^\pm \mu^\mp)&<&6.8\times 10^{-4} \, ,
\end{eqnarray}
which are significantly larger than those obtained in the renormalizable setting (see Eqs.~(\ref{BRHiggs-bounds1}) and
(\ref{BRHiggs-bounds2})). In particular, the $\tau^\pm \mu^\mp$ channel is remarkable from the experimental point of view, as it may be at the reach of future experiments.

As far as the $Z$ gauge boson decays are concerned, their branching ratios are bounded by
\begin{eqnarray}
BR(Z\to \mu^\pm e^\mp)&<&0.7\times 10^{-12} \ , \\
BR(Z\to \tau^\pm \mu^\mp)&<&1.1\times 10^{-7} \  .
\end{eqnarray}
These branching ratios are more constrained in comparison with those involving the Higgs boson, this is a direct consequence of the three orders of magnitude by which the $Z$ decay width exceeds the $ H $ decay width.

It is interesting to estimate the energy scales above which LFV, emerging from Lorentz violation, can arise. Let $Y^{AB}_0\equiv 1/\Lambda^{AB}_{LFV}$ be the new physics scale. From Eqs. (\ref{B1},\ref{B2}), the following bounds are derived:
\begin{eqnarray}
\Lambda^{\mu e}_{LFV}&>&8.3 \times 10^4 \, \, \, TeV \, ,\\
\Lambda^{\tau \mu}_{LFV}&>&204 \, \, \, TeV \, .
\end{eqnarray}
It is worth comparing these scales to the Fermi scale ($v  $) through the ratio $\epsilon^{AB}_{LFV}\equiv v/\Lambda^{AB}_{LFV}$. Using the above results, one obtains
\begin{eqnarray}
\epsilon^{\mu e}_{LFV}&=&3.0\times 10^{-6}\, , \\
\epsilon^{\tau \mu}_{LFV}&=&1.2\times 10^{-3}\, .
\end{eqnarray}
From these bounds, it is concluded that if the phenomenon of LFV is induced by LV, it occurs at very high energies, in particular, those transitions between the second and the first families.

Our results suggest that LFV may be detected at the LHC or ILC via the decay $H\to \tau^\pm \mu^\mp$. This possibility could have cosmological implications because of neutrinoless double beta decay, lepton number violating collider processes and the Baryon Asymmetry of the Universe seem to be tightly related~\cite{CI}.

To conclude this section, it is well worth comparing our results for the decay $H\to \tau^\pm \mu^\mp$ with those obtained within the context of an effective Lagrangian that preserve Lorentz invariance reported in Ref.~\cite{DT}. In this reference, this branching ratio was predicted using various dimension-six $SU_L(2)\times U_Y(1)$-invariant operators. Some of these operators predict, in certain scenarios, branching ratios of the same order of those obtained here. We will focus on the results implied by the operators given in Eq.(5) in Ref.~\cite{DT}. These operators generate generate a $H\tau^\pm \mu^\mp$ vertex similar to that induced in the nonrenomalizable version of the SME, that is, in both approaches this coupling is proportional to the $H$ derivative and then produced branching ratios of similar orders. Since in Ref.~\cite{DT}, the contribution of the operators appearing in Eqs.(5) was marginally studied, we present here some basic results in order to perform our comparison. The invariant amplitude is given in this case by
\begin{equation}
{\cal M}(H\to \tau \mu)=-\frac{\epsilon^{\tau \mu}}{v}[\bar{v}(p_2,s_2)\pFMSlash{p} u(p_1,s_1)]\, ,
\end{equation}
where $\epsilon^{\tau \mu}=(v/\Lambda^{\tau \mu}_{LFV})^2 \alpha^{\tau \mu}$, with $\alpha^{\tau \mu}$ a dimensionless constant. This amplitude must be compared with our result given by Eq.(\ref{MHiggs}), with $m_HY^{\tau \mu}_0$ playing the role of $\epsilon^{\tau \mu}$. Apparently, both amplitudes would lead to similar results for $\epsilon^{\tau \mu}$ and $m_HY^{\tau \mu}_0$ as being of the same order of magnitude. However, there is a subtlety directly associated with the fact that the $H$ momentum $p^\mu$ is contracted with the Dirac's matrix $\gamma_\mu$ in the Lorentz preserving case, whereas in the Lorentz violating case, this contraction occurs with the background field $Y^{\tau \mu}_\mu$. This fact manifests itself in the Dirac's traces that emerge once the amplitudes are squared. In fact, in the Lorentz preserving case, one has
\begin{equation}\fl
\qquad |{\cal M}(H\to \tau \mu)|^2=4\, m^2_\tau \, \left(\frac{m_H}{v}\right)^2\left(1-\frac{m^2_\mu}{m^2_\tau}\right)\left[1-\left(\frac{m_\tau +m_\mu }{m_H} \right)^2\right]|\epsilon^{\tau \mu}|^2\, ,
\end{equation}
which strongly depend on the mass squared difference $m^2_\tau -m^2_\mu$. On the other hand, in the Lorentz violating case the result is
\begin{equation}
|{\cal M}(H\to \tau \mu)|^2=2\, m^2_H \, \left[1-\left(\frac{m_\tau +m_\mu }{m_H} \right)^2\right]|m_HY^{\tau \mu}_0|^2\, ,
\end{equation}
which, in contrast, is proportional to $m^2_H$. This analysis shows that, for comparable values of $|\epsilon^{\tau \mu}|^2$ and $|m_HY^{\tau \mu}_0|^2$, the branching ratio for the $H\to \tau^\pm \mu^\mp$ decay in the Lorentz violating case would be more important than the Lorentz preserving one by about a factor of $(m_\tau/v)^2/2\approx 30258$. This is indeed the case, as the bounds for the $|\epsilon^{\tau \mu}|^2$ and $|m_HY^{\tau \mu}_0|^2$ factors obtained from the three-body decay $\tau \to \mu \bar{\mu}\mu$ do not differ significantly. In fact, the corresponding branching ratio in the Lorentz preserving context can be written as follows:
\begin{equation}
BR(\tau \to \mu \bar{\mu}\mu)=\frac{5\alpha^2 \left(g^{\tau \, 2}_V+g^{\tau \, 2}_A\right)}{768 s^4_{2W}}\left(\frac{m_\tau}{\Gamma_\tau}\right)\left(\frac{m_\tau}{m_Z}\right)^4|\epsilon^{\tau \mu}|^2\, ,
\end{equation}
where $\Gamma_\tau$ is the $\tau$ decay width and a convenient combination of the three $\epsilon$ parameters arising from the operators given by Eqs.(5) of Ref.~\cite{DT} has been used. In addition, the mass $m_\mu$ has been disregarded against  $m_\tau$. Notice the strong similitude between this expression and the one obtained in Eq.(\ref{TBD}). In fact, using updated experimental limits on this three-body decay one obtains a bound of  $|\epsilon^{\tau \mu}|^2<3\times 10^{-8}$, which is one order of magnitude more stringent than the one obtained for the Lorentz violating case [see Eq.(\ref{B2})]. A direct calculation shows that this updated bound leads to $BR(H\to \tau^\pm \mu^\mp)\sim 10^{-9}$ in this Lorentz preserving case.

\section{Conclusions}
\label{C}It is possible that the electromagnetic interaction exactly preserves the leptonic flavor, as it is suggested by the current experimental limits on photon-mediated transitions among charged leptons. However, lepton flavor violation could occur in nature via Higgs boson decays; at the end, this particle has a special role in the generation of the mass spectrum on the SM. The mass is an important quantum number which is correlated with the flavor of a particle, \textit{i.e.}, it is an intrinsic property directly associated with the identity of any elementary particle. The Higgs boson is the only particle in the SM that can distinguish the flavor through the mass because it couples to any massive particle with a coupling proportional to the mass itself. Lepton flavor violation may also be induced by the $Z$ gauge boson, the current experimental constraints on such decays are more flexible than those existing for the electromagnetic processes.

In this work, we have presented the analysis for lepton flavor violation mediated by the Higgs boson, the $Z$ gauge boson, and the photon in the context of an SME viewed as an effective field theory that incorporates non-conserved $CPT$  and Lorentz violation in a model-independent manner. We separately explored the aforementioned processes within renormalizable and nonrenormalizable extensions of  the SM Yukawa sector. In the renormalizable extension, the lepton flavor violation is induced at tree level by the three neutral bosonic particles of the SM; it was found that the presence of LFV induced in this way is quite suppressed due to severe constraints arising from experimental limits on the electromagnetic transitions $l_A\to l_B \gamma$. In this context, the Higgs decays into channels $\mu^\pm e^\mp$ and $\tau^\pm \mu^\mp$ have branching ratios of the order of $10^{-19}$ and $10^{-11}$, respectively. About the $Z$ boson decays into these channels, the respective branching ratios are of order of $10^{-21}$ and $10^{-12}$. These practically unobservable branching ratios are the consequence of the almost prohibited electromagnetic transitions  $l_A\to l_B \gamma$, generated at tree level in this renormalizable extension of the SM. On the other hand, in the nonrenormalizable extension of the Yukawa sector, lepton flavor violation is generated at tree level via both the Higgs and the $Z$ gauge bosons. In this scenario, lepton flavor violation mediated by the photon is naturally suppressed as this effect contributes at one-loop level. In this context, the Higgs decays into the channels $l^\pm_A l^\mp_B$ are proportional to the time component of the vector $Y^{AB}_\mu$ that characterizes the flavor violating transitions. Assuming a time-like $Y^{AB}_\mu$ vector, simple expressions for the $Z$ decays into  $l^\pm_A l^\mp_B$ were also derived. Under this assumption on $Y^{AB}_\mu$, the tree-level contribution of a virtual $Z$ gauge boson to the three-body decay $l_A \to l_B \bar{l}_C l_C$ was calculated. Then, the experimental constraints on these decays were used to bound the $Y^{AB}_\mu$ vector and thus allows us to predict the branching ratios for the Higgs and $Z$ bosons decays into the $\mu^\pm e^\mp$  and $\tau^\pm \mu^\mp$ channels. It was found that $BR(H\to \mu^\pm e^\mp)<4.1\times 10^{-9}$ and $BR(H\to \tau^\pm \mu^\mp)<6.8\times 10^{-4}$, whereas $BR(Z\to \mu^\pm e^\mp)<7.0\times 10^{-13}$ and $BR(Z\to \tau^\pm \mu^\mp)<1.1\times 10^{-7}$. The scales characterizing this class of new physics effects are $\Lambda^{\mu e}_{LFV}>8.3 \times 10^4$ TeV, for transitions between the second and first families, and $\Lambda^{\tau \mu}_{LFV}>204$ TeV, for transitions from the third to the second family. It can be seen from these results that, despite of relatively high new physics energy scale relative to the Fermi scale, the Higgs boson decay $H\to \tau^\pm \mu^\mp$ becomes compelling, as it can reach a branching ratio of almost $10^{-3}$ and thus be within the range of future measurements.

\section*{We acknowledge financial support from CONACYT (M\' exico) and SNI (M\' exico).}

\end{document}